\documentclass[conference]{IEEEtran}
\IEEEoverridecommandlockouts
\usepackage{cite}
\usepackage{amsmath,amssymb,amsfonts}
\usepackage{algorithmic}
\usepackage{graphicx}
\usepackage{textcomp}
\usepackage{xcolor}
\usepackage{hyperref}
\usepackage{multirow}
\usepackage{caption}
\usepackage{subcaption}
\hypersetup{
    colorlinks=true,
    linkcolor=blue,
    filecolor=magenta,      
    urlcolor=cyan,
    pdftitle={Property Graphs in Arachne},
    pdfpagemode=FullScreen,
    }
\usepackage[linesnumbered,ruled]{algorithm2e}
\def\BibTeX{{\rm B\kern-.05em{\sc i\kern-.025em b}\kern-.08em
    T\kern-.1667em\lower.7ex\hbox{E}\kern-.125emX}}
\begin{document}

\title{Property Graphs in Arachne}

\author{\IEEEauthorblockN{Oliver Alvarado Rodriguez, Fernando Vera Buschmann, Zhihui Du, David A. Bader}
\IEEEauthorblockA{\textit{Department of Data Science} \\
\textit{New Jersey Institute of Technology}\\
Newark, NJ, USA \\
\texttt{\{oaa9,fv54,zd4,bader\}@njit.edu}}
}

\maketitle

\begin{abstract}
Analyzing large-scale graphs poses challenges due to their increasing size and the demand for interactive and user-friendly analytics tools. These graphs arise from various domains, including cybersecurity, social sciences, health sciences, and network sciences, where networks can represent interactions between humans, neurons in the brain, or malicious flows in a network. Exploring these large graphs is crucial for revealing hidden structures and metrics that are not easily computable without parallel computing. Currently, Python users can leverage the open-source Arkouda framework to efficiently execute Pandas and NumPy-related tasks on thousands of cores. To address large-scale graph analysis, Arachne, an extension to Arkouda, enables easy transformation of Arkouda dataframes into graphs. This paper proposes and evaluates three distributable data structures for property graphs, implemented in Chapel, that are integrated into Arachne. Enriching Arachne with support for property graphs will empower data scientists to extend their analysis to new problem domains. Property graphs present additional complexities, requiring efficient storage for extra information on vertices and edges, such as labels, relationships, and properties.
\end{abstract}


\section{Introduction}
Property graphs are widely used in graph database systems to combine graph structures with attributes such as vertex labels, edge relationships, and properties. Data scientists often analyze networks that naturally store these attributes on vertices and edges. These attributes can enhance algorithms for tasks like breadth-first search on specific vertices or filtering subgraphs based on attribute matching, thereby enriching the data scientists' ability to analyze and understand the graph. It is essential to provide solutions for storing property graphs to enable data scientists to leverage the computational power of their systems effectively. These solutions should be integrated into proven libraries and frameworks designed for large-scale analysis, such as Arkouda \cite{merrill2019arkouda}. 

\looseness=-1
Arkouda is an open-source framework initially developed as a scalable replacement for NumPy in Python. Powered by Chapel  \cite{chamberlain2007parallel, chamberlain2011chapel, chamberlain2018chapel} at the backend and offering a Python interface, Arkouda has demonstrated its ability to handle datasets comprising over 500 million rows, making it an excellent choice for parallel analysis on large-scale datasets. With a user-friendly interface inspired by NumPy, Arkouda provides predefined operations for users to manipulate their datasets from Python scripts or Jupyter Notebooks. These operations primarily work with \textbf{p}arallel and \textbf{d}istributed array objects called \textbf{pd}arrays. Arkouda facilitates data preparation, exploration, and efficient parallel kernel invocation within a single session. Given that a significant amount of datasets can be structured as graphs, Arachne, built as an extension to Arkouda, facilitates efficient massive-scale graph analysis \cite{2022-RDPLB}.

Arachne aims to be a highly productive graph framework for data scientists looking to extract information efficiently from large graph datasets. It introduces a distributable graph data structure called the Double-Index data structure (DI) \cite{du2021interactive}. Arachne includes implementations of graph kernels such as breadth-first search and triangle counting, which can be executed on both shared-memory and distributed-memory systems. This work focuses on enhancing Arachne's analysis capabilities by introducing additional structures to enhance DI for property graphs.

The main contributions in this paper are as follows:
\begin{enumerate}
    \item \textbf{DIP}, a data structure derived from the \textbf{DI} data structure, specifically designed to store \textbf{p}roperty graphs.
    \item Various versions of DIP implemented in Chapel, exploring space and time-efficient variations: DIP-LIST, DIP-LISTD, and DIP-ARR with experimential results.
\end{enumerate}
All of our results are reproducible based off functionality found at https://github.com/Bears-R-Us/arkouda-njit for property graph analysis.

\section{The Property Graph Data Model}
A property graph is a directed and labeled multigraph composed of a set of vertices $V$ and edges $E$. Each vertex $v \in V$ and edge $(u,v) \in E$ can store property key-value pairs. Vertices store labels and edges store relationships, where each edge between two vertices with a distinct relationship is considered its own unique edge \cite{angles_property_2018}. If an edge has multiple relationships this means there is a multiedge, i.e., multiple copies of one edge. 
\looseness=-1

\begin{figure*}[htbp]
    \centerline{\includegraphics[width=0.75\textwidth]{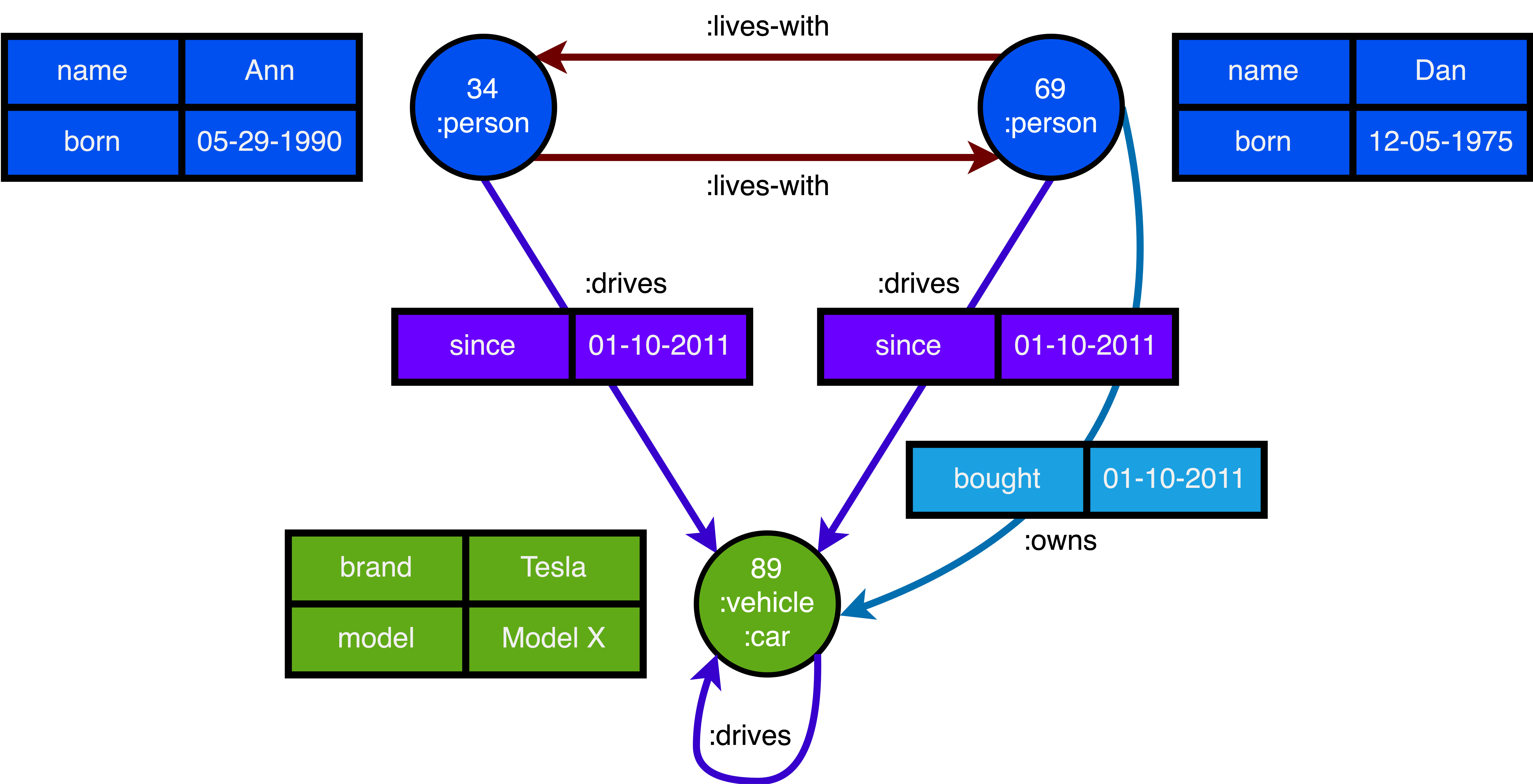}}
    \caption{Example of a property graph with three vertices and five edges (the two edges between vertices with values 69 and 89 are structurally maintained as one but can conceptually be considered two distinct edges). The tables show the properties that are defined on each vertex as well as some of the edges. The label, relationship, and property sets can be empty as is the case with \texttt{lives-with}.}
    \label{fig:property_graph}
\end{figure*}

Property graphs can be either static or dynamic. In static property graphs, edges and/or vertices cannot be added into the graph over time, whereas dynamic graphs allow for the addition of edges and vertices over time. For this paper, we only target static property graphs built from datasets that can be viewed as dataframes where vertex labels, edge relationships, and properties can all be inferred from the columns of a tabular dataset. An example of a property graph can be seen in Fig. \ref{fig:property_graph}.
\looseness=-1

Given two vertices $u,v \in V$ and an edge $e \in E$ where $e = (u,v)$, then it is said that the source vertex is $u$ and the destination vertex is $v$ where there is a direction specified as $u \rightarrow v$. Data can be extracted from the property graph data model when given some vertex $u$ or edge $e=(u,v)$. These operations can be thought of as queries on the data structure where the information stored at these locations is returned back to the user upon completion.
\looseness=-1

\section{DI Fundamentals}
\label{sec:di_fund}
DI was first introduced into Arachne by Du \emph{et al.} \cite{du2021interactive} to allow for easy distribution of edges across a compute cluster. In this section, we will highlight the fundamentals of DI for directed graphs. DI is typically composed of four arrays: source ($SRC$), destination ($DST$), number of neighbors ($NEI$), and the starting indices ($STR$) into $SRC$ and $DST$. For this work, we optimized its space complexity by amalgamating the $STR$ and $NEI$ arrays into one array called $SEG$. The grouping of the $SRC$ and $DST$ arrays are referred to as the edge index arrays, whereas $SEG$ is referred to as the vertex index array. The indices of the edge arrays are in the range $[0,m-1]$ and the indices of the vertex array are in the range $[0,n]$ where $m=|E|$ is the number of edges and $n=|V|$ is the number of vertices. For the $SEG$ array the first index is always $0$ and the end index is always $m$. Given an edge $e=(u,v)$, the vertices are stored in the edge arrays where $SRC[e]=u$ and $DST[e]=v$ and $e$ is the index into the edge arrays. All the edges in $SRC$ and $DST$ are sorted based off the vertex values where $SRC$ is sorted first, and then for every vertex, its corresponding adjacency list is sorted in $DST$. The vertex index array is created based off the sorted edge arrays. Lastly, all the original vertex names are normalized to the range $[0,n-1]$ during construction. Storing these arrays takes $\Theta(m) + \Theta(n)$ space.

\begin{figure}[htbp]
    \centerline{\includegraphics[width=0.35\textwidth]{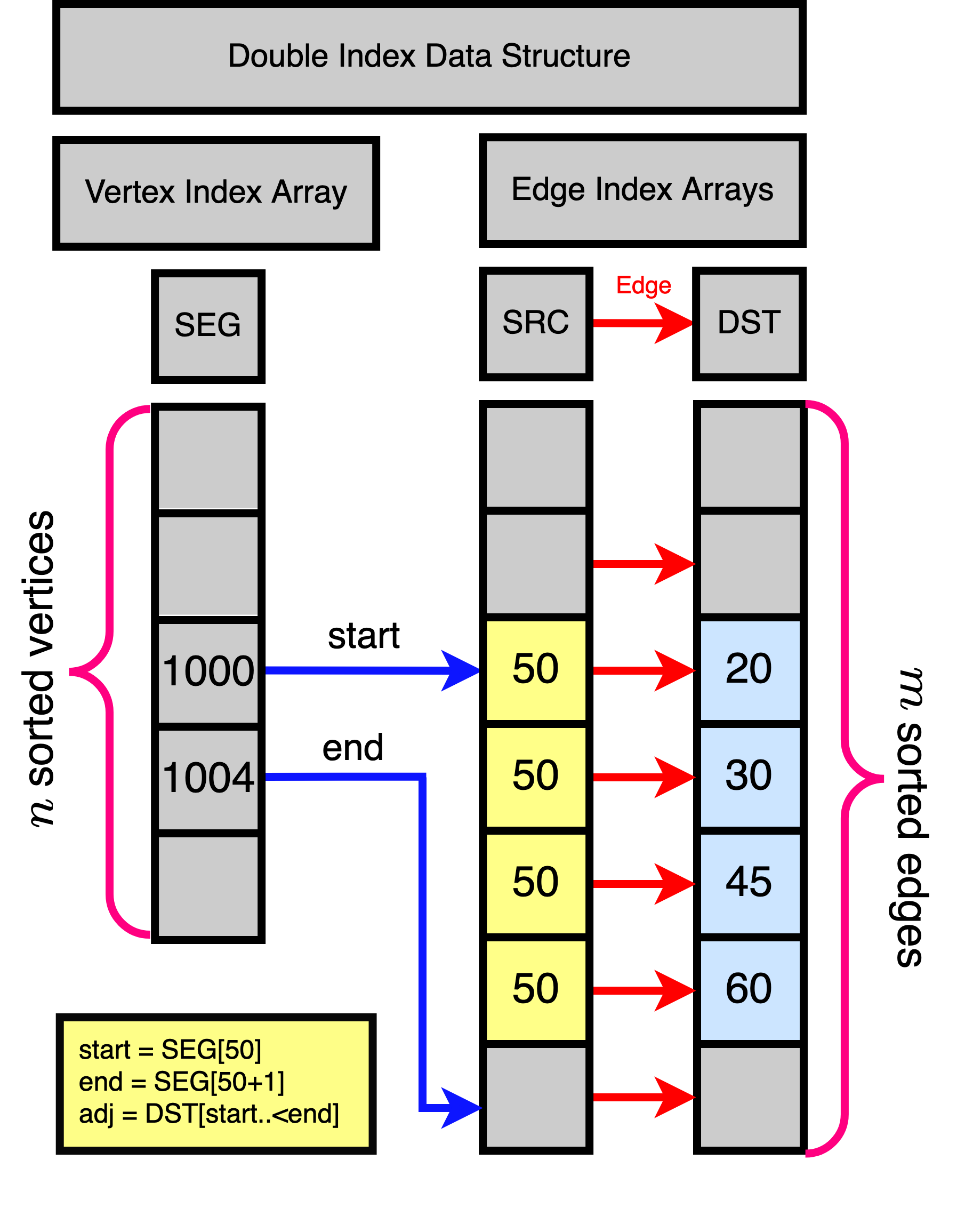}}
    \caption{Example of neighborhood slicing in DI. To get the neighborhood of the vertex with index $50$, the slice is taken of $DST[SEG[u]..SEG[u+1]-1] = DST[1000..1003]$. The number of neighbors for $u$ can be taken by using $SEG[u+1] - SEG[u]$. The $SEG$ domain set is the range $[0,n]$ and the domain set range of $SRC$ and $DST$ is $[0,m-1]$. The domain map specified by those domain sets makes up the indices of those arrays.}
    \label{fig:slice_example}
\end{figure}

Given a vertex identifier $u$, the neighborhood of that vertex $u$ can be found by using the following Chapel array slice $DST[SEG[u]..SEG[u+1]-1]$. The edge and vertex index arrays are distributed in a block-distributed manner to the compute nodes that are allocated for the job. In Chapel, the result of an array slice is a reference to the subset of the array elements specified from the slicing index set. No new memory is ever allocated, making this operation memory efficient. An example showing the slicing can be found in Fig. \ref{fig:slice_example}. If a vertex $u$ has $k$ neighbors then the time to iterate over the adjacency list is $\Theta(k)$ and finding this list takes constant time $O(1)$. DI enhances CSR by explicitly listing all edges to facilitate both edge-based and vertex-based algorithms. 

\section{DIP Design and Development}
DIP is powered by the DI data structure that currently drives graph storage in Arachne. It employs the same edge-centric view of graphs that allows for easy load-balancing across cluster (multilocale) systems in Chapel. Since DIP is designed to be written in Chapel, we will discuss operations in terms of how they are implemented in Chapel.

\subsection{Notes on the DIP Design}
Everything listed in Sec. \ref{sec:di_fund} is applicable to DIP with the added complexity of storing multiple vertex labels, edge relationships, and properties. While designing DIP and its variations, we approached the problem in a memory-efficient manner to ensure we also matched the compactness of DI. We implemented three different methods to store property graphs based off of two-dimensional byte arrays (DIP-ARR), Chapel domains (DIP-LIST), and Chapel lists (DIP-LISTD). Vertices and edges are referred to as entities whereas their labels, relationships, and properties are referred to as attributes. In short, attributes are either represented by a two-dimensional byte array that flags whether a particular entity contains it, or by lists that maintain a single copy of every attribute for each entity.

\subsection{DIP-LIST(D)}
\label{sec:DIP-ll}
Storing attributes can be done in an attribute-centric manner where we store each attribute for a vertex or edge explicitly, and for the case of DIP-LISTD we maintain pointers to the ``next" and ``previous" attribute to easily extract all the vertices and/or edges that make up that attribute. This is the typical method used in many graph databases where objects represent each vertex and store all the data held by that vertex. This choice, however, is not very memory efficient as each storing object must maintain pointers and entity/attribute identifiers. 

\begin{figure}[htbp]
    \centerline{\includegraphics[width=0.3\textwidth]{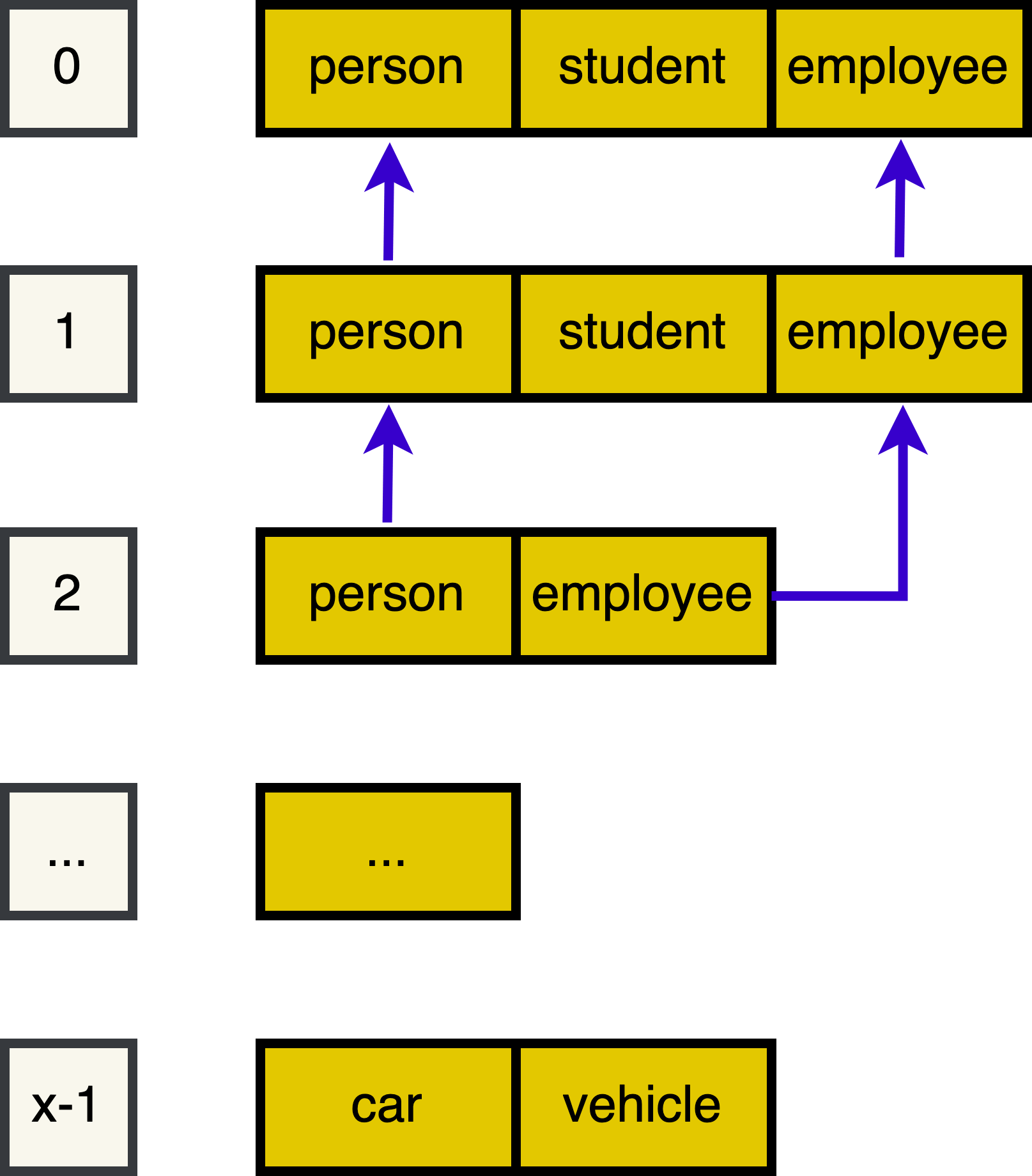}}
    \caption{Example of storing attributes in lists for each element. In this case, there are $x$ entities where $x$ can be either $n$ or $m$ depending on whether vertices or edges are being stored. In the purple arrows, we show how DIP-LISTD maintains an extra way of searching the data structure backwards to only traverse the entities that make one particular attribute.}
    \label{fig:dip_ll}
\end{figure}

\looseness=-1
An example of both DIP-LIST and DIP-LISTD can be seen in Fig. \ref{fig:dip_ll}. For the case of DIP-LIST the list stored for a particular entity just contains an integer representing the input string. On the other hand, for DIP-LISTD we store \texttt{Node} objects that contain variables to store the data, vertex or edge they belong to, and pointers to the next and previous elements that induce a doubly-linked list. Since this is a distributable data structure, the previous and next pointers point to objects that can live on the different locales allocated for a job. Objects in Chapel are pointers to heap-allocated memory. Using the Chapel memory management technique called \texttt{shared}, an object can be initialized and allocated at runtime and remain in scope fully until all variables that reference that object go out of scope. Changes to the data are allowed and done through the memory management technique \texttt{borrowed} which does not delete the object when the borrowing variable goes out of scope. 


The addition of a \texttt{Node} during property graph construction requires updating the \texttt{next} pointer of the previous \texttt{Node} and the \texttt{prev} pointer of the \texttt{Node} being inserted. This is done by extracting the last added \texttt{Node} from \texttt{last\_entity\_tracker}. Once that is finished, \texttt{last\_entity\_tracker} is updated to include the \texttt{Node} that was just added. For this map, the key is the name of the attribute and the value is the \texttt{Node} that was just added. The addition requires calling a lock on both the map and list until the insertion operation finishes. This is done by encapsulating the code running the operation with a mutex lock created by using \texttt{sync} variables in Chapel. Currently, this is not as efficient as it could be, and optimizing these insertions are left to future work.

\subsection{DIP-ARR}
\label{sec:dip_arr}
\begin{figure}[htbp]
    \centerline{\includegraphics[width=0.5\textwidth]{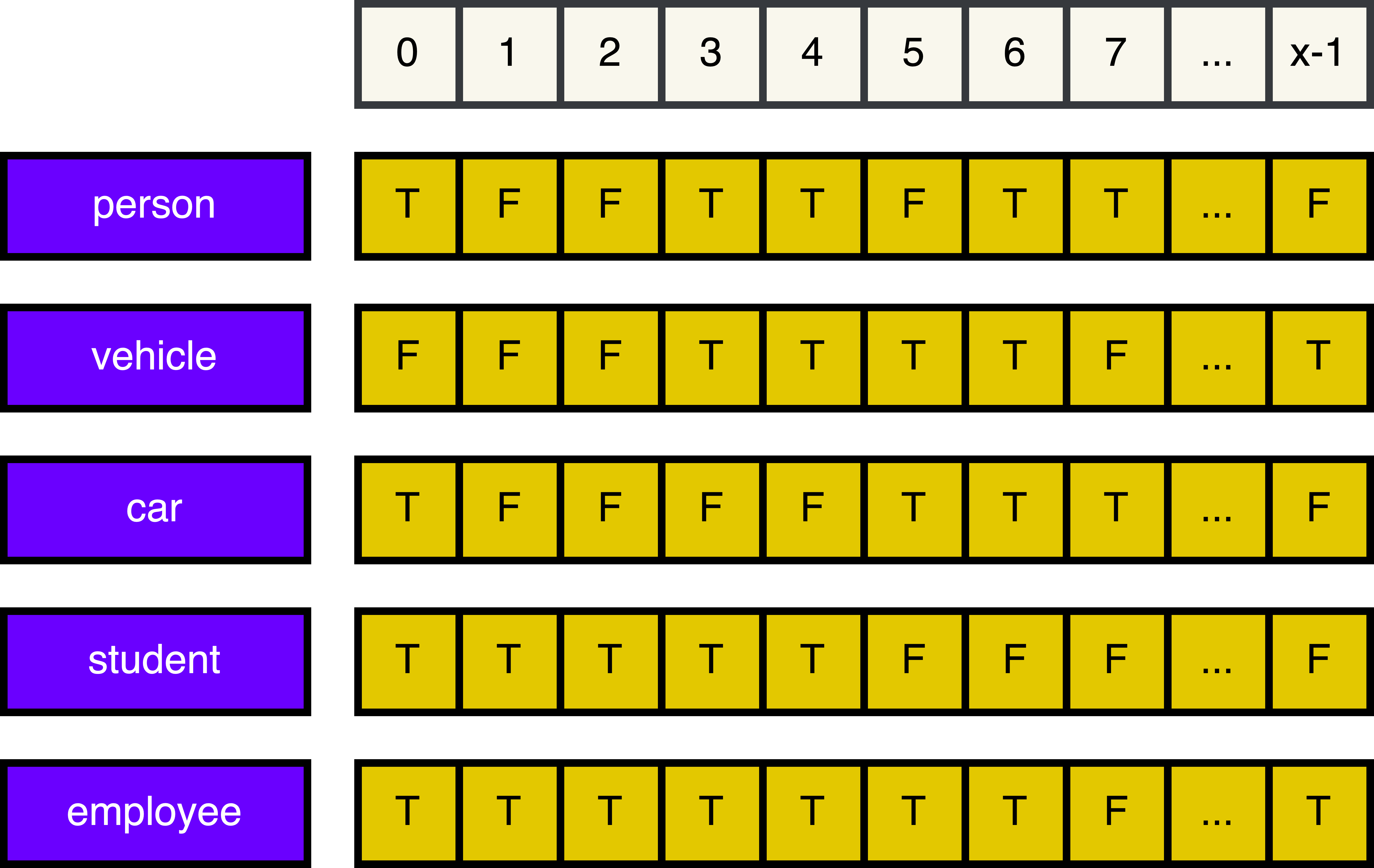}}
    \caption{Example of storing attributes as a two-dimensional byte Boolean array. The number of columns is of size $x$ which is either $n$ or $m$ depending on whether vertices or edges are being stored. The number of attributes stored can be of any size $k$, in this case $k=5$. To extract the value stored for a given vertex or edge, if it is true, the row integer identifier is passed through a sorted array to return the original value of the string.} 
    \label{fig:dip_array}
\end{figure}

Unlike the list versions of DIP that were implemented, we also implemented an array-based data structure that makes indexing and slicing for data more efficient and avoids complicated class structures to represent the data stored. Further, traversing arrays is relatively inexpensive since they are stored contiguously in memory. In this section, we explore our array-based method of storing attributes. Simply put, for each attribute there exists a Boolean array of size $n$ or $m$ depending on whether it is storing vertex or edge information. Then, storing that specific attribute is just storing \texttt{true} if it exists for an entity and \texttt{false} otherwise. An example of this can be seen in Fig. \ref{fig:dip_array}. 

The two-dimensional Boolean byte array is partitioned into chunks using the array type \texttt{domain(2) dmapped Block({0..<k, 0..<x})} in Chapel. This operation creates a blocked array with two dimensions with $k$ rows and $x$ columns. It is chunked in such a way by Chapel that if there were four locales then the array would be split into four quadrants, one for each locale. This would mean that no one entire attribute list for an entity or entity list for an attribute would be stored on the same machine. However, this should not impact performance much during querying processes since each locale only processes the array chunk it owns. 

\subsection{Space and Time Complexity Trade-offs}
\label{sec:complexity}
Each of the proposed variations of DIP supports the same fundamental operations of insertion and querying. To reiterate, vertex and edges are referred to as entities and labels and relationships as attributes. The insertion operations are specified for inserting attributes. The querying operations are specified for returning all the attributes specified for an entity, or accepting a list of attributes and returning all the entities that contain any of them. The returned values can be further processed to find the intersections of the returned vertex and edge arrays to create a subgraph. We will use $N$ to refer to the size of the entity set and $K$ to refer to the size of the attribute set. We will use $k \leq K$ to denote the size of an attribute set for any given entity. 

\subsubsection{Space Complexity}
\label{sec:space_complexity}
DIP-LIST stores a list for each entity of size $k$ that varies for each entity. In the worst case, each entity will contain every attribute to make the size of DIP-LIST to be $O(NK)$. DIP-LISTD stores the same list with extra data that has constant -- but not neglible -- size. This constant $c$ makes the storage of DIP-LISTD to be, in the worst case, $O(cNK)$. This $c$ will be made up of 64B for the attribute integer id, 64B for each vertex id (which is doubled for edges), 8B for the previous pointer, and 8B for the next pointer. This creates a total of 208B for edge attributes and 144B for vertex attributes. Lastly, DIP-ARR stores a two-dimensional array of size $N \times K$ making its space complexity $\Theta(NK)$.


\subsubsection{Building Time Complexity}
DIP-LIST and DIP-LISTD insert data sequentially led by parallel chunks of work. This means that we can populate two vertices $u$ and $v$ that live on separate chunks simultaneously, but changes to the domains or lists for $u$ and $v$ must be done sequentially to avoid race conditions. This comes out to a time of $O(\frac{cNK}{P})$ where $N$ is the number of entities, $c$ is the overhead of inserting into a list or domain, $K$ is the number of attributes being inserted, and $P$ is the number of processors. In the case of DIP-ARR, we set a flag if we encounter that attribute for an entity. Thus, this time complexity is $O(\frac{NK}{P})$, where $P$ is the number of parallel processing units in the system.

\section{Data Ingestion Workflow}
\label{sec:diworkflow}
Currently, Arachne targets the same data science workflows targeted by Arkouda. Therefore, it is assumed that property graphs are generated from data already in-memory that has been read in by Arkouda from file formats such as HDF5, Parquet, or CSV files. Arachne contains the capability to read in matrix market files, but the ability to store vertex and edge attributes in this format is limited. Therefore, the time it takes to preprocess these datasets with Arkouda is not taken into consideration here, and all workflows are assumed to begin after the original data ingestion and cleansing.

When the data are already present in Arkouda, the base data structure (DI) is constructed by Arachne from two Arkouda arrays that signify the source and destination vertices of an edge. This is achieved by creating a graph with Arachne using the property graph class, \texttt{graph = ar.PropGraph()} and adding edges in bulk to it through \texttt{graph.add\_edges\_from(source, destination)}. Once the graph is populated with vertices and edges, graph attributes follow. It is expected that the data scientist will load attributes independently from different dataframes they generate from their data. Four independent functions are available to handle each of the four types of attributes.

Typically, the steps to ingest property graph data involves three main steps. (1) Remap attribute values to an integer identifier to reduce storage space. (2) Generate internal indices of vertices and edges that correspond to where data will be stored in the back-end. (3) Insert the data into DIP in the back-end. Steps 1 and 2 are facilitated by existing Arkouda functionality and step 3 is written in Chapel at the back-end. Manipulating array-based data is highly efficient in Arkouda which Arachne exploits to increase performance.

\section{Querying Data}
The property graph data model allows us to search for entities or attributes that match a particular query. These queries specified on property graphs can follow different formats \cite{francis_cypher_2018}, but all queries boil down to simple searches on the graph data structure. Creating a data structure that allows fast and easy searching with parallel reads will increase performance as one increases the number of processors that the system runs on. Fast querying makes data analysis more interactive and improves data science workflow uptime. We will follow the same notation and worst-case scenarios as specified in Sec. \ref{sec:complexity}. For querying in this paper, we define it as passing a string array with any number of attributes and returning the entities that contain them. When an entity is found with any of the passed attributes, they are marked as true and the final array returned is a Boolean array that marks which entity indices make up the returned query.

\subsection{DIP-LIST}
Given an attribute, finding all the entities that contain it takes $O(\frac{NK}{P})$ time since every single attribute list for every entity must be traversed. The fraction $\frac{N}{P}$ breaks the data up into blocks where each search is done sequentially by the task spawned to tackle that block.

\subsection{DIP-LISTD}
Given an attribute, finding all the entities that contain it takes $O(N)$ time since we traverse starting from the last \texttt{Node} added into \texttt{last\_entity\_tracker} (see Sec. \ref{sec:DIP-ll}). This traversal involves parsing through previous and next pointers in the distributed memory doubly-linked list. Since Chapel objects are just pointers to a distributed heap-allocated space, jumping to an object stored on a different locale requires spawning a thread on the remote locale to process that object.

\subsection{DIP-ARR}
Given an attribute, finding all the entities that contain it takes $O(\frac{N}{P})$ time since we traverse the row for the given attribute to see which elements are \texttt{true}. This method is the simplest to parallelize since Chapel tasks run concurrently on the locale that owns a slice of the array.

\section{Experiments}
Experiments were conducted by varying a configuration of 1, 2, and 8 compute nodes (locales). Each locale consists of 128 cores (64 per AMD EPYC 7713 CPUs), 1TB DDR4 RAM, and an Infiniband HDR 200 GB/s node interconnect. Further, the number of cores per locale utilized varied between 32, 64, and 128 cores. This variance is done due to the fact that Chapel runs a single process per locale and then uses multiple threads per locale for concurrency. Each of those threads can be issuing remote communications which goes through GASNet. Communication injection is serialized within GASNet for Infiniband networks, therefore increasing the number of cores can degrade performance for codes that perform a large amount of fine-grained communication. For graph building and adding in attributes, decreasing the number of cores degraded performance, but not significantly. However, querying was heavily improved by reducing the number of cores due to the current nature of the code performing many fine-grained communications when writing the entities that match the query. Therefore, we limit our results to show scalability as the number of locales are increased when setting the number of cores to 32. Large-scale experiments are delegated to future work. Here, we show a simple scalability measure of our methods. 

\subsection{Datasets}
Graphs were generated randomly by creating two arrays of a given size (number of edges) and populating them with random integers from a given range. For these experiments, we set the random vertex integers created to be that of the same size as number of edges to minimize the amount of multiple edges that are created. Graph information is given in Tab. \ref{tab:graphs}. For this experimental study, the structure of the graph is not taken into consideration nor how it can impact execution times. In other words, inspecting the graph for regularity or power-law distributions is left to future work. We increase the number of edges for each graph by 10x. The set sizes for the number of labels and relationships was set to 50 and the vertices and edges populated with labels were randomly selected from a pool equal to the vertex and edge sets. Some vertices or edges could be repeated and some not selected at all.

\begin{table*}[thbp]
\centering
\caption{Information for randomly generated graphs. The number of vertices ($n$), number of edges ($m$), minimum in/out-degrees, maximum in/out-degrees, and average in/out-degrees are all shown.}
\begin{tabular}{|r|r|r|r|r|r|r|r|r|}
\hline
\multicolumn{1}{|l|}{} & \textbf{$n$} & \textbf{$m$}    & \textbf{min in-deg} & \textbf{max in-deg} & \textbf{avg in-deg} & \textbf{min out-deg} & \textbf{max out-deg} & \textbf{avg out-deg} \\ \hline
\textbf{graph1}        & 86,503     & 100,000       & 0                   & 8                   & 1                   & 0                    & 9                    & 1                    \\ \hline
\textbf{graph2}        & 864,237    & 1,000,000     & 0                   & 10                  & 1                   & 0                    & 8                    & 1                    \\ \hline
\textbf{graph3}        & 8,646,309  & 10,000,000    & 0                   & 9                   & 1                   & 0                    & 10                   & 1                    \\ \hline
\textbf{graph4}        & 86,469,224 & 100,000,000   & 0                   & 11                  & 1                   & 0                    & 11                   & 1                    \\ \hline
\textbf{graph5}        & 864,648,454  & 1,000,000,000 & 0                   & 12                  & 1                   & 0                    & 11                   & 1                    \\ \hline
\end{tabular}
\label{tab:graphs}
\end{table*}

\begin{figure}[htbp]
    \centerline{\includegraphics[width=0.4\textwidth]{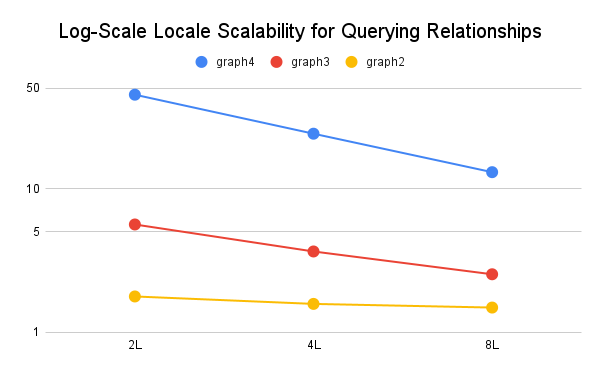}}
    \caption{Log-scale scalability of execution times as the number of locales is increased for DIP-LIST. There is a visible downward trend for graphs 3 and 4, with less visibility for graph 2 due to its small size.}
    \label{fig:res1}
\end{figure}

\begin{figure}[htbp]
    \centerline{\includegraphics[width=0.4\textwidth]{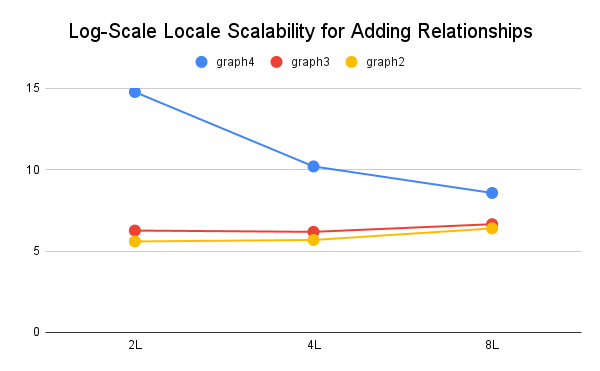}}
    \caption{Log-scale scalability of execution times as the number of locales is increased for DIP-LIST. There is a visible downward trend for graph 4 with an upward curve for graphs 2 and 3.}
    \label{fig:res2}
\end{figure}

\subsection{Results}
\looseness=-1
We now highlight results for graph building and ingesting/querying relationships. Due to the fact that adding any type of attribute requires the same basic steps as highlighted in Sec. \ref{sec:diworkflow}, for DIP-LIST we show results on relationship operations. We omit results for DIP-LISTD because across the board its operations were up to 10x slower than DIP-LIST and DIP-ARR. 

For querying operations on both DIP-LIST and DIP-ARR, the execution time dropped as the number of locales were increased from 2 to 4 to 8. We can see these results for DIP-LIST in Fig. \ref{fig:res1}. DIP-ARR also followed a downward trend but it was not as drastic as the drops we see for DIP-LIST, therefore we omit those results from this section. The reason for this performance increase is because no traversals are being performed between locales. Quite simply, the more locales that are added, the more resources are available for each of them to independently process their chunk of the property graph stored. For adding relationships, the trend is more apparent for graph 4 as seen in Fig. \ref{fig:res2}.

The largest graph tested was graph 5 from Tab. \ref{tab:graphs} on eight locales. Adding relationships to it took 30.43 seconds and querying its relationships took 118.38 seconds, less than two minutes to entirely return the edge set of a new graph that matched the query space. This translates to 8.5 million edges processed per second for query operations. For adding labels and relationships, the most time consuming operations were the remapping of vertex values and index generation steps. The actual internal storage of values amounted to less than three seconds for graph5 meaning that built-in Chapel data structures such as \texttt{domains} are highly efficient.

\section{Related Work}
Property graphs concentrate on the labels, properties, and relationships of vertices and edges and how they can be used to increase the knowledge extracted from them \cite{7840959,BALES2006451}. The work by McColl \emph{et al.} \cite{mccoll_performance_2014} provides a performance evaluation of open-source graph databases, where most store their data using the property graph data model. The simplest way to store graph-based data models is via a labeled property graph, which is a set of triples. The work by Angles \emph{et al.} \cite{angles_multilayer_2022} provides a new way of viewing graph-based data called multilayer graphs that extends directed labeled graphs with edge identifiers.

Property graphs utilize a graphical representation, where vertices represent entities and edges represent the relationships between them. This graphical approach provides a visual representation of the data structure. Property graphs allow for representing connections between entities and the properties associated with vertices and edges \cite{angles_rdf_2019, EhrlingerW16}. This capability enables the storage and querying of detailed information about the entities and their relationships within the graph. Property graphs are employed for data analysis and discovering hidden knowledge \cite{7095609,7840959}. These models support advanced queries and analysis, facilitating the extraction of meaningful information from the graph. Property graphs focus on representing the properties and relationships of vertices and edges in a graph\cite{angles_rdf_2019}. Property graphs model data using vertices, edges, and properties without the need for a predefined schema \cite{angles_rdf_2019}. Property graphs offer more flexibility in terms of adding new properties or relationships between vertices and edges, allowing for quicker adaptation to changes in data requirements\cite{10.1145/3567444}. Property graphs employ database-specific query languages such as Cypher in the case of Neo4j\cite{8257957}.

\section{Conclusion}
Designing data structures for property graphs involves not only efficiently storing the vertices and edges of a graph, but more importantly, the attributes are also stored with them. Oftentimes, property graph database developers want to tightly couple data with the entity, as was shown in the DIP-LIST and DIP-LISTD data structures. DIP-LIST and DIP-ARR allow for fast traversals and storing large amounts of data on multiple locales easily, and efficiently. Further work involves optimizing the DIP-LIST method that allows for easy label and relationship additions with fast querying. Further, this work can be easily extended for property storage and algorithms that utilize property graphs.

\section*{Acknowledgment}
We thank the Chapel and Arkouda communities for their guidance. This research is supported in part by the NSF grant CCF-2109988.

\bibliographystyle{plain}
\bibliography{arkouda-chapel,suffixarray,Property_Graphs,importgraph,graph,KG}

\end{document}